\documentclass{article}

\PassOptionsToPackage{numbers, compress}{natbib}
\usepackage[final]{nips_2017}
\usepackage{bm}
\usepackage[utf8]{inputenc} 
\usepackage[T1]{fontenc}    
\usepackage{hyperref}       
\usepackage{url}            
\usepackage{booktabs}       
\usepackage{amsfonts}       
\usepackage{nicefrac}       
\usepackage{microtype}      
\usepackage{graphicx}
\usepackage{verbatim} 
\usepackage{color}
\usepackage{amsmath,amssymb,graphicx}

\title{Shape optimization in laminar flow with a label-guided variational autoencoder}

\author{
  Stephan~Eismann \\
  Stanford University\\
  \texttt{seismann@cs.stanford.edu} \\
  \And
  Stefan~Bartzsch \\
  TU Munich \\
  \texttt{stefan.bartzsch@tum.de} \\
  \And
  Stefano~Ermon \\
  Stanford University \\
  \texttt{ermon@cs.stanford.edu} \\
}

\begin{document}

\maketitle

\begin{abstract}
Computational design optimization in fluid dynamics usually requires to solve non-linear partial differential equations numerically. In this work, we explore a Bayesian optimization approach to minimize an object's drag coefficient in laminar flow based on predicting drag directly from the object shape. Jointly training an architecture combining a variational autoencoder mapping shapes to latent representations and Gaussian process regression allows us to generate improved shapes in the two dimensional case we consider.
\end{abstract}

\section{Introduction}

\label{introduction}
The optimization of hydro- and aerodynamic properties for vehicles, turbines and engines is a common and challenging engineering problem. The design process usually involves both numeric simulations as well as experimental measurements. Due to the large complexity of the design space success often relies on experience and skill of the engineer \cite{kumar_future_2000}. 

In this work we explore whether Bayesian optimization can facilitate the design process. 
As an exemplary and general problem we discuss the resistance of an object in a fluid (e.g. air or water). What shape should a vesicle have to reduce energy dissipation by drag to a minimum? The resistance of an object in fluid dynamics is characterized by the drag coefficient $c_d$ \cite{landau_fluid_2013}, 
\begin{equation}
c_d = \frac{2 F_d}{\rho v^2 A}.
\label{eqn:cd}
\end{equation}
In the reference frame of the object, $F_d$ is the component of the force caused by the fluid in flow direction which represents drag. $\rho$ is the density of the fluid, $v$ the flow velocity and $A$ the frontal area of the object in flow direction. $c_d$ is not a constant but depends on velocity, viscosity and other parameters which are summarized in the Reynolds number $Re$, a dimensionless quantity. Together $c_d$ and $Re$ which are mostly empirically determined govern the drag of an object.  \\

The Navier-Stokes equations provide a theoretical description of the flow of a fluid around an object \cite{landau_fluid_2013}, 
\begin{equation}
\rho \frac{\partial \bm{v}}{\partial t} + \rho (\bm{v} \cdot \nabla) \bm{v} = -\nabla p + \eta \nabla^2\bm{v},
\label{eqn:NSE}
\end{equation}
where $p$ denotes the pressure and $\eta$ the viscosity coefficient. Equation \ref{eqn:NSE} describes a second order non-linear partial differential equation and is except for a few special cases only numerically solvable. Here we consider the laminar (i.e. non-turbulent) flow of an incompressible fluid that obeys
\begin{equation}
\nabla \cdot \bm{v} = 0,
\end{equation}
in the limit of low Reynolds numbers $Re$. The drag forces on the object have two components: (i) friction drag which is caused by the viscosity of the fluid and (ii) pressure drag which is caused by the pressure distribution around the object's surface.   

Finally, it should be noted that Equation \ref{eqn:cd} strictly speaking only holds true for a limited range of $Re$. In the domain of laminar flow at low $Re$, the drag of an object is proportional to the velocity $v$ (Stokes law) \cite{landau_fluid_2013}. Drag only becomes proportional to the square of velocity when turbulences occur in the wake field of the object at larger velocities. Nevertheless, $c_d$ provides a quantitative account of the drag in laminar flow if $v$ is kept constant throughout all simulations. 

Previous works have demonstrated the successful application of kriging or Gaussian process (GP) regression for the optimization of parametric designs in aerodynamic applications \cite{simpson_kriging_2001,martin_use_2005, jeong_efficient_2005,jouhaud_surrogate-model_2007}. Here, we present a purely data-based approach. The combination of variational autoencoder (VAE) and Gaussian process allows us to learn and optimize shape-characteristics in a non-parametric form directly from images.  

\section{Data and model}
\label{data_model}

\subsection{Laminar flow in 2D}
We consider the case of laminar flow around an object in two dimensions as it allows us to generate training and test data at relatively low computational cost. 

\begin{figure}[h]
  \centering
  \includegraphics[width=0.8\textwidth]{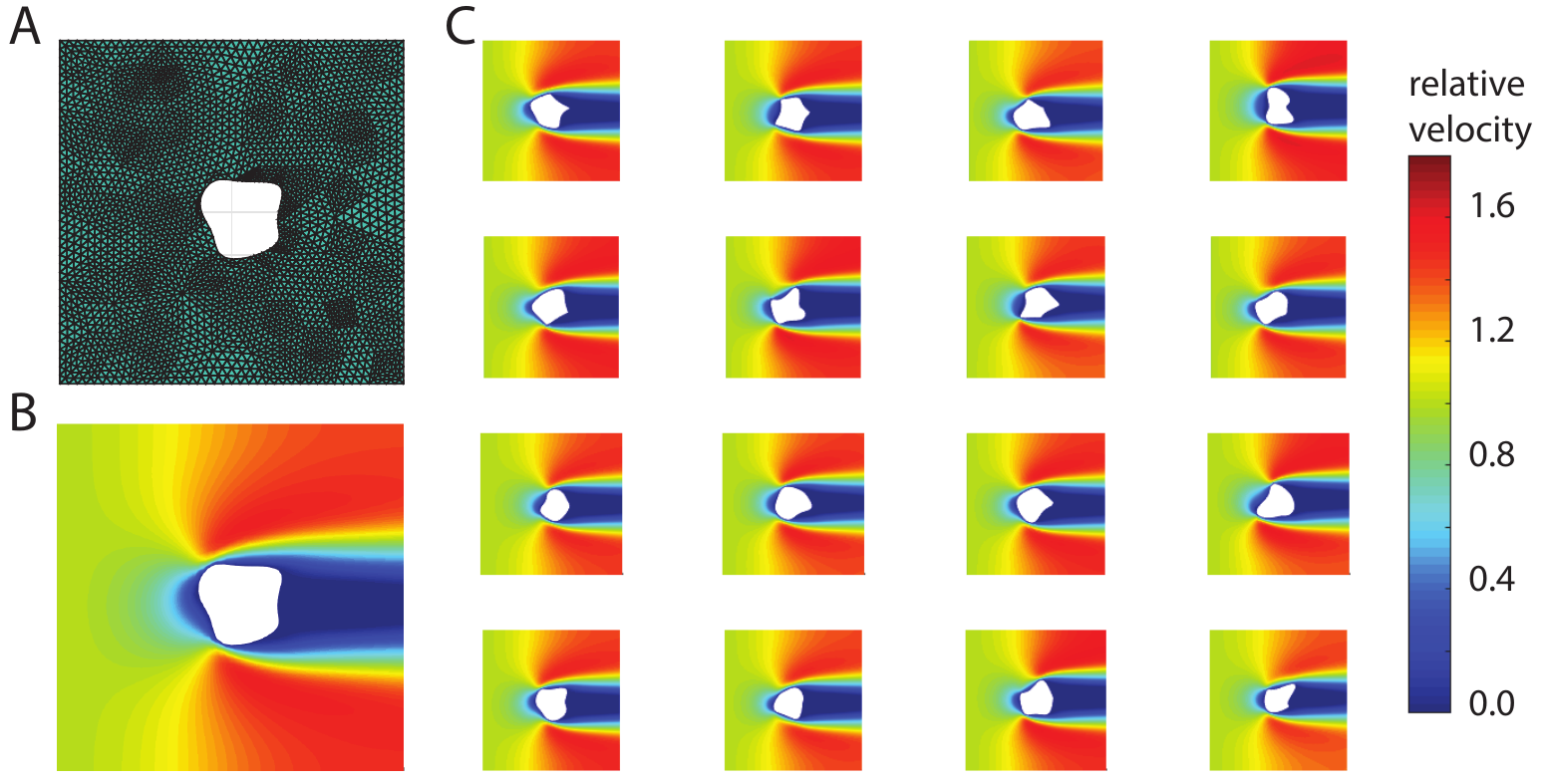}
  \caption{\textbf{Generation of training data}  
  (\textbf{A}) Meshed geometry around a randomly generated shape. (\textbf{B}) Color-coded fluid velocity around the shape calculated as a solution to the Navier-Stokes equations. The left wall defines an inlet for a flow in positive x-direction. Red and green indicate regions of high and low flow velocity, respectively. (\textbf{C}) Color-coded velocity fields for a range of randomly generated shapes.}
\label{fig:data_generation}
\end{figure}

Figure \ref{fig:data_generation} illustrates the simulation setup and computation. We generate random shapes by picking radii from a uniform distribution at different polar angles. Subsequently we use a set of Fourier descriptors to obtain a smooth shape which facilitates the geometry meshing (Fig.\ref{fig:data_generation}A). The discretized space allows us to solve the Navier-Stokes equations (Eq.\ref{eqn:NSE}) with a standard finite element solver (QuickerSim, MATLAB). The left wall is defined as fluid inlet with constant flow profile directed in positive x-direction. Furthermore we apply slip boundary conditions to the upper and lower wall (i.e. $v_y = 0$) and consider slip at the object boundaries. Figures \ref{fig:data_generation}B,C show the resulting fluid velocity fields for a range of shapes. In analogy to the three dimensional case we define the frontal area A as the maximum object extension in y-direction as projected onto the x-axis. All simulations are performed with a scalar viscosity $\nu =0.02$. Generating a dataset of 5000 shapes requires about $16$h on a single machine (Intel i7-3520M with 2.90 GHz, 16GB memory). 

\subsection{Guided shape encoding and Bayesian optimization}
\begin{figure}[h]
  \centering
  \includegraphics[width=1.0\textwidth]{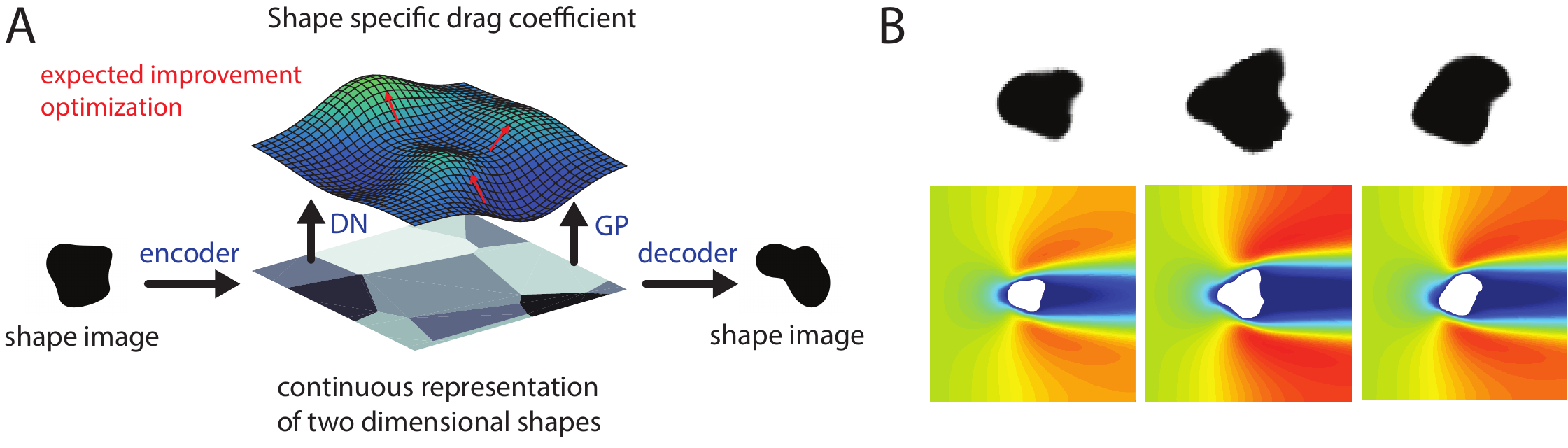}
  \caption{\textbf{Model architecture and drag optimized shapes} 
  (\textbf{A}) Images of object shapes are mapped to and from a continuous latent space by an encoder-decoder pair of neural networks. An additional neural network (DN) maps the latent space feature vectors to drag coefficient. Combining the networks' loss functions yields a simple way to label-guided encoding. To obtain an uncertainty measure for the predictions we also train a sparse Gaussian process (GP) and generate new shapes by optimizing expected improvement (EI) through gradient ascent.  (\textbf{B}) Three shapes corresponding to local EI maxima and the corresponding velocity-field solutions. The generated shapes are slightly fuzzy as is characteristic for VAEs. All shapes improve upon the best example in the training set with respect to drag coefficient.}
\label{fig:scheme-shapes}
\end{figure}

Figure \ref{fig:scheme-shapes}A shows a schematic of the model architecture. The architecture is similar to the approach described in \citet{gomez-bombarelli_automatic_2016} for the design of drug-like molecules and light-emitting diodes.

Object shapes are encoded as binary images of size $112 \times 84$ and mapped to and from a continuous latent space $\mathbf{z}$ of dimensionality $\mathrm{d}$ by two neural networks forming the encoder and decoder pair of a variational autoencoder (VAE) \cite{kingma_auto-encoding_2013}. The encoder consists of two convolutional and two dense layers. The decoder is designed similarly with two dense followed by two deconvolutional layers. 

We map the latent feature vectors $\mathbf{z}$ to drag coefficient $y$ through a sparse-spectrum Gaussian process (GP) regression model \cite{lazaro-gredilla_sparse_2010}. We choose the sparse approach to avoid the computational complexity of a full GP. Next to the most probable drag coefficient, the Gaussian process provides us also with an uncertainty estimate when predicting the drag of unobserved points in the latent space. The probabilistic measure allows us to calculate expected improvement (EI) \cite{jones_efficient_1998} for each new point with respect to the smallest drag coefficient observed so far and optimize towards larger EI values using gradient ascent. Finally, we decode latent feature vectors associated with large EI values to obtain corresponding object shapes.

In the approach of \citet{gomez-bombarelli_automatic_2016} VAE and GP are trained separately and sequentially. Here we introduce an additional neural network termed `drag network' (DN) mapping from latent space $\mathbf{z}$ to drag coefficient $y$ and whose parametric form allows us to enforce label-guided encoding in a straightforward manner. We define the joint loss function $\textrm{loss}_\mathrm{JOINT}$ as the sum of VAE loss (reconstruction plus KL divergence) and the mean-squared regression error of the DN: 
\begin{equation}
\label{eq:joint}
 \textrm{loss}_\mathrm{JOINT} = \textrm{loss}_\mathrm{VAE} + \textrm{loss}_\mathrm{DN} \quad .
\end{equation}
The $\textrm{loss}_\mathrm{DN}$ term encourages the encoder to find a representation that is amenable for regression. However, DN does not provide a measure of uncertainty for its predictions, hence it is not suitable for shape optimization. We therefore additionally train the GP regression model on the latent space, keeping the encoder fixed.

\section{Results}
\subsection{Comparing joint and separate training}
We consider a dataset of 5200 shapes with corresponding drag coefficients and randomly assign $80\%$ and $20\%$ of the data to training and test set, respectively. Table \ref{tab:results} compares two different training modalities for the models: (i) VAE and drag network are trained jointly (Equation \ref{eq:joint}), (ii) VAE and drag network are trained independently in a sequential manner. All models were trained using the same set of hyperparameters except for latent dimensionality and the number of units per dense layer which is doubled for columns $\left(2,4,6,8\right)$ compared to $\left(1,3,5,7\right)$. 

On average the separately trained variational autoencoders perform better in reconstructing shapes from the test and training set as expected. Reconstruction loss here is measured as the mean squared deviation per pixel. The same holds true for the predictive performance of the drag network on the training set. However, the situation is reversed in the test situation in which the jointly trained models do much better both in terms of Gaussian process and drag network prediction. 

\begin{table}[h]
\centering
\begin{tabular}{lcccc|cccc}
    \toprule
 & \multicolumn{4}{c}{\textbf{Joint training}} & \multicolumn{4}{c}{\textbf{Separate training}} \\
 \midrule
Latent dimensions & 10 & 10 & 20 & 20 & 10 & 10 & 20 & 20  \\
GP: MSE test & 0.64	& 0.77 & 0.64 & \textbf{0.52} & 0.76 & 0.76 & 1.00 & 0.74  \\
DN: MSE test & \textbf{0.54}	& 0.61 & 0.56 &	\textbf{0.54} & 0.92	& 0.87 & 1.00 &	0.87 \\
DN: MSE train & 1.00 &	0.72 & 0.94 & 0.91 & 0.54 &	\textbf{0.48} & 0.58 & 0.47 \\
Reconstr. test & 0.60 &	1.00 & 0.40 & 0.38	& 0.50	& 0.46 & 0.95 & \textbf{0.28}  \\
Reconstr. train & 0.67 & 0.80 & 0.43 & 0.35 & 0.56 & 0.51 & 1.00 & \textbf{0.30} \\
\bottomrule 
\\
\end{tabular}
\caption{\textbf{Guided encoding enhances predictive performance for test shapes} \newline
The table compares test and training errors for models of varying latent dimensionality and joint and separate training of VAE and DN. Errors are relative to the largest error among all models. The number of units per dense layer is doubled for experiments in columns $\left[2,4,6,8\right]$ compared to $\left[1,3,5,7\right]$. For these models the DN performed better on the test set than the sparse spectral GP which we attribute to the relative small number of inducing points (500) and a lack of their optimization with respect to marginal likelihood. Best results are shown in bold.}
\label{tab:results}
\end{table}

\subsection{Optimizing for drag-reduced shapes}
After training an additional Gaussian process model (1000 inducing spectral points, 20 latent dimensions, joint training) on the dataset mentioned above we generate new object shapes by optimizing expected improvement with respect to the smallest drag coefficient in our training data. To do so we randomly select 2000 starting points from the latent distribution of the training set and perform parallel gradient ascent. The latent points corresponding to the largest local maxima are then decoded to new object shapes. Overall training requires about 7h (TensorFlow implementation on a NVIDIA TITAN Xp) with the expected improvement optimization performed locally (Intel i7-3520M with 2.90 GHz, 16GB memory) requiring an additional 1h.   \\

Fig. \ref{fig:scheme-shapes}B shows the shapes and computed velocity fields for the three best out of a total of 25 decoded latent points after one round of Bayesian optimization. The decoded shapes are slightly blurry as is characteristic for VAE-generated images. We perform Sobel edge detection to extract a set of boundary points for the shapes and can subsequently calculate the corresponding drag coefficient. All three shapes slightly outperform the best training data point with improvements in the range of $4-8\%$. 

\section{Discussion}
Computational design in fluid dynamics is costly due to the requirement of numerically solving non-linear differential equations. Here we present a data-driven Bayesian approach to design optimization. We are able to generate object shapes with an improved drag coefficient compared to the training dataset in the case of two dimensional, laminar flow. In addition, we demonstrate the benefit of label-guided encoding for shape-based drag prediction. Natural extensions of this work include the application of the presented approach to the case of three dimensions and turbulent flow. 

\small
\bibliography{flow_nips17}

\end{document}